# Stationary Bound States of Dirac Particles in Collapsar Fields

M.V. Gorbatenko, V.P. Neznamov[1]

RFNC-VNIIEF, 37 Mira Ave., Sarov, 607188, Russia

Abstract

Stationary bound states of elementary spin ½ particles that do not decay with time are obtained for a Schwarzschild gravitational field using a self-conjugate Hamiltonian with a flat scalar product at small values of gravitational coupling constant.

In order to obtain a discrete energy spectrum, we introduce a boundary condition such that the azimuthal current density of Dirac particles on the "event horizon" is zero.

The results can lead to revisiting some concepts of the standard cosmological model related to the evolution of the universe and interaction of collapsars with surrounding matter.

---

[1] E-mail: neznamov@vniief.ru

The classical Schwarzschild solution is characterized by a point spherically symmetrical source of gravitational field of mass $M$ and "event horizon" (gravitational radius).

$$r_0 = \frac{2GM}{c^2}. \quad (1)$$

In (1), $G$ is the gravitational constant, and $c$ is the speed of light. For a particle of mass $m$, the dimensionless gravitational coupling constant equals

$$\alpha = \frac{GMm}{\hbar c} = \frac{Mm}{m_p^2} = \frac{r_0}{2l_c}. \quad (2)$$

In (2), $\hbar$ is the Planck constant, $m_p = \sqrt{\frac{\hbar c}{G}} = 2.2 \cdot 10^{-5}$ gram is the Planck mass, and $l_c = \frac{\hbar}{mc}$ is the Compton wavelength.

For an electron, $\alpha \simeq 1$ corresponds to $M = 0,5 \cdot 10^{15}$ $kg$. The value of the coupling constant corresponding to a source having the mass of the Sun ($M = M_\odot \approx 2 \cdot 10^{30}$ $kg$) for an electron-mass particle is $\alpha \cong 4 \cdot 10^{15}$.

Despite the evident electromagnetic analogy in atomic physics, bound states of Dirac particles in the Schwarzschild field have been investigated comparatively scantily. For the gravitational case, there is a belief that bound states have complex energies. In this case, these states decay exponentially with time. The existence of resonant Schwarzschild states for massive scalar particles using the Klein-Gordon equation was discussed in [1]-[2]. The same problem for massive Dirac particles was discussed in [3]-[9]. In these papers, a hydrogen-like spectrum with relativistic corrections was obtained for $\alpha \ll 1$ by direct solution of the Dirac equation in a weak Schwarzschild field for the real part of energy.

In papers [10]-[12], we develop a method for deriving self-conjugate Dirac Hamiltonians with a flat scalar product within the framework of pseudo-Hermitian quantum mechanics for arbitrary, including time dependent, gravitational fields. Apparently, such stationary Hamiltonians, if there are square integrable wave functions and if corresponding boundary conditions are specified, will provide existence of stationary bound states of Dirac particles with a real energy spectrum. We suppose that for these cases Hamiltonians are Hermitian ($(\Psi, H\varphi) = (H\Psi, \varphi)$).

This paper is devoted to defining stationary bound states of a Dirac particle in a Schwarzschild field.

Below we will use the system of units $G = \hbar = c = 1$, signature

$$\eta_{\underline{\alpha}\underline{\beta}} = \text{diag}[1, -1, -1, -1] \quad (3)$$

and notations: $f = 1 - \dfrac{r_0}{r}$; $\gamma^0$, $\gamma^k$, $k = 1,2,3$ are local 4x4 matrices in the Dirac-Pauli representation.

Then, the Schwarzschild metric in the $(t,r,\theta,\varphi)$ coordinates will take the form:

$$ds^2 = f \cdot dt^2 - \dfrac{dr^2}{f} - r^2 \left( d\theta^2 + \sin^2\theta d\varphi^2 \right). \tag{4}$$

The self-conjugate Hamiltonian $H_\eta$ corresponding to this metric equals [12]

$$\begin{aligned} H_\eta = &\sqrt{f} m \gamma^0 - i\sqrt{f} \gamma^0 \Big\{ \gamma^1 \sqrt{f} \left( \dfrac{\partial}{\partial r} + \dfrac{1}{r} \right) + \\ &+ \gamma^2 \dfrac{1}{r}\left( \dfrac{\partial}{\partial \theta} + \dfrac{1}{2}\mathrm{ctg}\,\theta \right) + \gamma^3 \dfrac{1}{r\sin\theta} \dfrac{\partial}{\partial \varphi} \Big\} - \dfrac{i}{2}\dfrac{\partial f}{\partial r}\gamma^0 \gamma^1. \end{aligned} \tag{5}$$

In (5), we imply real values of $f > 0$.

The Dirac equation for stationary states $\psi(r,\theta,\varphi,t) = e^{-iEt}\psi(r,\theta,\varphi)$ takes the form:

$$\begin{aligned} \gamma^0 E \psi(r,\theta,\varphi) = &\Big\{ \sqrt{f} m - i\gamma^1 \left[ f\left( \dfrac{\partial}{\partial r} + \dfrac{1}{r} \right) + \dfrac{r_0}{2r^2} \right] - \\ &- i\sqrt{f}\left[ \gamma^2 \dfrac{1}{r}\left( \dfrac{\partial}{\partial \theta} + \dfrac{1}{2}\mathrm{ctg}\,\theta \right) + \gamma^3 \dfrac{1}{r\cdot\sin\theta}\dfrac{\partial}{\partial \varphi} \right] \Big\} \psi(r,\theta,\varphi). \end{aligned} \tag{6}$$

As we know, Eq. (6) allows for separation of variables, if the bispinor $\psi(r,\theta,\varphi)$ is given by

$$\psi(r,\theta,\varphi) = \begin{pmatrix} F(r)\ \xi(\theta) \\ -iG(r)\ \sigma^3 \xi(\theta) \end{pmatrix} e^{im_\varphi \varphi} \tag{7}$$

and the following equation is used (see e.g. [13])

$$\left[ -\sigma^2 \left( \dfrac{\partial}{\partial \theta} + \dfrac{1}{2}\mathrm{ctg}\,\theta \right) + i\sigma^1 m_\varphi \dfrac{1}{\sin\theta} \right] \xi(\theta) = i\kappa \xi(\theta). \tag{8}$$

In order to receive Eq. (8) we made an equivalent replacement of matrices in Eq. (6):

$$\gamma^1 \to \gamma^3, \gamma^3 \to \gamma^2, \gamma^2 \to \gamma^1.$$

In the equalities (7), (8): $\xi(\theta)$ are spherical harmonics for spin 1/2, $\sigma^i$ are two-dimensional Pauli matrices, $m_\varphi$ is the magnetic quantum number, $\kappa$ is the quantum number of the Dirac equation:

$$\kappa = \pm 1, \pm 2 ... = \begin{cases} -(l+1),\ j = l + \dfrac{1}{2} \\ l,\ \ \ j = l - \dfrac{1}{2} \end{cases}. \tag{9}$$

In (9) $j, l$ are the quantum numbers of the total and orbital momentum of a Dirac particle.

As a result, we obtain a system of equations for real radial functions $F(r), G(r)$. Let us write it in dimensionless variables $\varepsilon = \dfrac{E}{m}$, $\rho = \dfrac{r}{l_c}$, $\dfrac{r_0}{l_c} = 2\alpha$.

$$\left(1 - \frac{2\alpha}{\rho}\right)\frac{\partial F}{\partial \rho} + \left(\frac{1 + \kappa\sqrt{1 - \dfrac{2\alpha}{\rho}}}{\rho} - \frac{\alpha}{\rho^2}\right)F - \left(\varepsilon + \sqrt{1 - \frac{2\alpha}{\rho}}\right)G = 0,$$

$$\left(1 - \frac{2\alpha}{\rho}\right)\frac{\partial G}{\partial \rho} + \left(\frac{1 - \kappa\sqrt{1 - \dfrac{2\alpha}{\rho}}}{\rho} - \frac{\alpha}{\rho^2}\right)G + \left(\varepsilon - \sqrt{1 - \frac{2\alpha}{\rho}}\right)F = 0.$$

(10)

The range of variation of the variable $\rho$ for the functions $F(\rho), G(\rho)$ is $(2\alpha, \infty)$ in accordance with the metric (4) and presence of expression $\sqrt{1 - \dfrac{2\alpha}{\rho}}$ in Eqs. (10).

Eqs. (10) show that, as in the classical case, quantum mechanics prohibits the presence of Dirac particles under the "event horizon" $r \leq r_0$, i.e. at $\rho \leq 2\alpha$.

In order to construct a numerical solution, let us consider asymptotics of the system (10).

For $\rho \to \infty$ the leading terms of asymptotics equal

$$F = C_1 e^{-\rho\sqrt{1-\varepsilon^2}} + C_2 e^{\rho\sqrt{1-\varepsilon^2}}$$
$$G = \sqrt{\frac{1-\varepsilon}{1+\varepsilon}}\left(-C_1 e^{-\rho\sqrt{1-\varepsilon^2}} + C_2 e^{\rho\sqrt{1-\varepsilon^2}}\right).$$

(11)

In order to provide the finite particle motion, one should use only exponentially vanishing solutions (11), i.e. in this case $C_2 = 0$.

Since

$$\rho = 2\alpha \frac{r}{r_0},$$  (12)

with increase in $\alpha$, one should use increasingly higher values of $\rho$ in the numerical solutions of the system (10).

For $\rho \to 2\alpha$ $(r \to r_0)$,

$$F = \frac{A}{\sqrt{\rho - 2\alpha}}\sin\left(2\alpha\varepsilon\ln(\rho - 2\alpha) + \varphi\right),$$

$$G = \frac{A}{\sqrt{\rho - 2\alpha}}\cos\left(2\alpha\varepsilon\ln(\rho - 2\alpha) + \varphi\right),$$

(13)

where $A$ and $\varphi$ are constant values.

The oscillating functions $F$ and $G$ in (13) are ill-defined at the "event horizon", but they are quadratically integrable functions at $\rho \neq 2\alpha$

In order to find bound states in the system (10), one should define the boundary condition for $\rho \to 2\alpha$ $(r \to r_0)$.

The condition that there should be no particles under the horizon can be implemented if we suppose that components of the current density of Dirac particles for $\rho \to 2\alpha$ $(r \to r_0)$ are equal to zero.

For the Schwarzschild metric (4), components $j^r = \psi^+ \gamma^0 \gamma^3 \psi = 0$, $j^\theta = \psi^+ \gamma^0 \gamma^1 \psi = 0$ because of representation (7) and form of angle functions $\xi(\theta)$.

Azimuthal component

$$j^\varphi = \psi^+ \gamma^0 \gamma^2 \psi = 2F(\rho)G(\rho)\left[\xi^+(\theta)\sigma^1\xi(\theta)\right] \neq 0, \tag{14}$$

The boundary condition $j^\varphi \to 0$ at $\rho \to 2\alpha$ reduces to the condition

$$F(\rho)G(\rho)\big|_{\rho \to 2\alpha} \to 0. \tag{15}$$

From two possible variants of implementation (15), we will fulfil it using equality

$$G(\rho_N)\big|_{\rho \to 2\alpha} \to 0. \tag{16}$$

Some reason for this is known smallness of function $G(\rho)$ in comparison with function $F(\rho)$ in nonrelativistic approximation of Dirac equation.

Considering (7), (13), the condition (16) turns into

$$\begin{aligned} &\cos(2\alpha\varepsilon\ln(\rho-2\alpha)+\varphi) \to 0 \text{ when } \rho \to 2\alpha \\ &2\alpha\varepsilon\ln(\rho-2\alpha)+\varphi \to \frac{\pi}{2}N, \ N = \pm 1, \pm 3, \pm 5... \end{aligned} \tag{17}$$

Preliminary numerical simulations of the system (10)[2] were done to determine the stationary real energy spectrum of a Dirac particle of mass $m$ in the Schwarzschild field. They allow us to draw the following conclusions:

1. For $\alpha \ll 1$, the energy spectrum is nearly identical to the hydrogen-like spectrum

   $$E_n \approx m\left(1 - \frac{\alpha^2}{2n^2}\right).$$

2. Numerical relativistic corrections for the values of $\alpha = 0,05; 0,1$ are close to the corrections determined in [7] according to the formula

---

[2] Numerical implementation of the system (10) was provided by M.A. Vronsky. The methods for solving the system (10) will be presented in the next paper.

$$\frac{\Delta E_n}{mc^2} = -\frac{3\alpha^4}{n^4}\left\{\frac{n}{l+\frac{1}{2}}\left[\left(1-\frac{1}{3}\delta_{l0}\right)+\frac{1}{12}(1-\delta_{l0})\left(\frac{1}{\kappa}+\frac{2}{l(l+1)}\right)\right]-\frac{5}{8}\right\}. \quad (18)$$

A more detailed analysis of the numerical simulations and their results in a wide range of the gravitational coupling constant will be presented in the next paper. In our further studies, we are also going to present results of numerical simulations which define stationary bound states of Dirac particles in the Reissner-Nordström and Kerr fields.

The major result of this work is that we demonstrated the existence of stationary bound states of spin ½ Dirac particles in the Schwarzschild gravitational field when using the self-conjugate Hamiltonian (5) and boundary conditions (11), (17).

The results obtained are special in that the wave functions of particles down the horizon are equal to zero, and that collapse in its standard understanding is excluded for bound particles. Along with this, energy levels at $\alpha > 1$ can be fairly deep, so the binding energy can be on the order of $mc^2$. The results of this work can lead to revisiting some concepts of the standard cosmological model related to the evolution of the universe and interaction of collapsars with surrounding matter.

**Acknowledgement**

We thank our colleagues M.A. Vronsky and E.Yu. Popov, I.I.Safronov for useful discussions.

# References

[1] N.Deruelle, R.Ruffini. Phys. Lett., **52B**, 437, 1974.

[2] T.Damour, N.Deruelle, R.Ruffini. Lett. Nuov. Cim., **15**, 257, 1976.

[3] I.M.Ternov, V.P.Khalilov, G.A.Chizhov, A.B. Gaina. Proceedings of Soviet Higher Educational Institutions, Physics, No.9, 1978 (in Russian).

[4] A.B.Gaina, G.A.Chizhov. Proceedings of Soviet Higher Educational Institutions, Physics, No.4, 120, 1980 (in Russian).

[5] I.M.Ternov, A.B.Gaina, G.A.Chizhov. Sov. Phys. J. 23, 695-700, 1980.

[6] D.V.Galtsov, G.V.Pomerantseva, G.A.Chizhov. Sov. Phys. J. 26, 743 – 745, 1983.

[7] I.M.Ternov, A.B.Gaina. Sov. Phys. J., 31 (2); 157-163, 1988.

[8] A.B.Gaina, O.B.Zaslavskii. Class. Quantum Grav., 9; 667-676, 1992.

[9] A.B.Gaina, N.I.Ionescu-Pallas. Rom. J. Phys. 38, 729 – 730, 1993.

[10] M.V.Gorbatenko, V.P.Neznamov. Phys. Rev. D 82, 104056 (2010); arxiv: 1007.4631 (gr-qc).

[11] M.V.Gorbatenko, V.P.Neznamov. Phys. Rev. D 83, 105002 (2011); arxiv: 1102.4067v1 (gr-qc).

[12] M.V.Gorbatenko, V.P.Neznamov. Arxiv: 1102.0844v3 (gr-qc).

[13] D.R. Brill, J.A.Wheeler. Rev. of Modern Physics, 29 (1957), 465-479.